\documentclass[%
 reprint,
superscriptaddress,
 aps,
prl,
]{revtex4-2}

\usepackage{amsmath}
\usepackage{graphicx} 
\usepackage{subfig}
\usepackage{braket}
\usepackage{bbold}
\usepackage{amssymb}
\usepackage{mathrsfs}
\usepackage{dcolumn}
\usepackage{bm}
\usepackage{hyperref}
\usepackage{ragged2e} 

\newcommand{\bb}{\boldsymbol}
\newcommand{\mean}{\mathbb{E}}

\begin{document}

\preprint{APS/123-QED}

\title{Fundamental Limits on Clock Precision from Spacetime Uncertainty in Quantum Collapse Models
}

\author{Nicola Bortolotti}
 \email{Contact author: nicola.bortolotti@uniroma1.it}
\affiliation{Centro Ricerche Enrico Fermi - Museo Storico della Fisica e Centro Studi e Ricerche “Enrico Fermi”, Via Panisperna 89 A, 00184, Rome, Italy.}
\affiliation{Physics Department, “Sapienza” University of Rome, Piazzale Aldo Moro 5, 00185, Rome, Italy.}
\affiliation{Laboratori Nazionali di Frascati, Istituto Nazionale di Fisica Nucleare, Via Enrico Fermi 54, 00044, Frascati, Italy.}
\author{Catalina Curceanu}
\email{Contact author: catalina.curceanu@lnf.infn.it}
\affiliation{Laboratori Nazionali di Frascati, Istituto Nazionale di Fisica Nucleare, Via Enrico Fermi 54, 00044, Frascati, Italy.}
\affiliation{IFIN-HH, Institutul National pentru Fizica si Inginerie Nucleara Horia Hulubei, 30 Reactorului, 077125, M\u agurele, Romania.}
\author{Lajos Di\'osi}
\affiliation{Wigner Research Center for Physics, H-1525 Budapest 114 , P.O.Box 49, Hungary.}
\affiliation{E\"otv\"os Lor\'and University, H-1117 Budapest, P\'azm\'any P\'eter stny, 1/A, Hungary.}
\author{Simone Manti}
\affiliation{Laboratori Nazionali di Frascati, Istituto Nazionale di Fisica Nucleare, Via Enrico Fermi 54, 00044, Frascati, Italy.}
\author{Kristian Piscicchia}
\affiliation{Centro Ricerche Enrico Fermi - Museo Storico della Fisica e Centro Studi e Ricerche “Enrico Fermi”, Via Panisperna 89 A, 00184, Rome, Italy.}
\affiliation{Laboratori Nazionali di Frascati, Istituto Nazionale di Fisica Nucleare, Via Enrico Fermi 54, 00044, Frascati, Italy.}

\date{\today}

\begin{abstract}
Models of spontaneous wavefunction collapse explain the quantum-to-classical transition without invoking the von Neumann measurement postulate. Prominent frameworks, such as the Di\'osi-Penrose (DP) and Continuous Spontaneous Localization (CSL) models, propose a continuous, spontaneous measurement of the mass density field of quantized matter. We show that this mechanism could link both models - not just DP - to fundamental uncertainties in Newtonian gravity. Despite their non-relativistic nature, these models suggest an induced uncertainty in the flow of time due to fluctuations in the Newtonian potential. We calculate the ultimate limit on time uncertainty and demonstrate that the resulting clock-time uncertainty remains negligible for all contemporary time-keeping devices, including atomic clocks. 
\end{abstract}

\maketitle

\textit{Introduction}$-$Despite the remarkable experimental success of quantum mechanics, its standard formulation retains a somewhat ad hoc character. Within this framework, the quantum-to-classical transition is explained through quantum measurements that induce abrupt wavefunction collapses, requiring measurement devices and, in some interpretations, even a human observer \cite{wigner1995remarks, london1939theorie, stapp2004mind, chalmers2021consciousness}.

The debate over the correct interpretation of quantum mechanics has persisted throughout the last century and remains unresolved. Among the various proposals, Spontaneous Collapse Models (SCMs) - also referred to as Dynamical or Objective - offer a compelling alternative \cite{ghirardi1986unified, pearle1989combining, diosi1989models, diosi1987universal, ghirardi1990continuous, ghirardi1990markov, penrose1996gravity, bassi2013models}. These models not only provide a consistent interpretation of quantum mechanics that aligns with existing experimental observations, they also predict novel, testable effects that are currently under intense experimental investigation \cite{carlesso2022present}.

Models of spontaneous wavefunction collapse assume that quantum measurements occur continuously and everywhere in nature. The associated wavefunction collapses - often referred to as dynamical, spontaneous, or objective - provide an alternative explanation for the quantum-to-classical transition without invoking the standard quantum measurement postulate. The price to pay is the presence of a certain universal irreversibility that modifies the unitary dynamics of the wavefunction. 

This irreversibility may stem from the intrinsic uncertainty in the structure of spacetime \cite{karolyhazy1966gravitation, diosi2018fundamental}. Over the past decades, two non-relativistic proposals of SCMs, known as Diósi-Penrose (DP) model \cite{diosi1987universal, diosi1989models} and Continuous Spontaneous Localization (CSL) model \cite{ghirardi1990markov}, have attracted growing interest. Although these models share several key features, it is commonly believed that the nature of this uncertainty differs, as the DP model explicitly relies on the hypothesis of spacetime fluctuations, whereas CSL does not. Here we point out, for the first time, that CSL too can be related to spacetime uncertainties. 

The latency of CSL's gravity-relatedness is of reason. The DP model corresponds to the spontaneous measurement of the mass density field, which assumes the uncertainty of the Newtonian field \cite{diosi1987universal, diosi1989models} and even generates it \cite{tilloy2016sourcing, tilloy2017principle}. In contrast, CSL originated from the Ghirardi-Rimini-Weber (GRW) model \cite{ghirardi1986unified}, which posits that each particle in the universe undergoes spontaneous localization processes, randomly in time and independently of one another. The original version of CSL involved the spontaneous measurement of the number density field, occurring continuously in both space and time, in accordance with the GRW framework. Eventually, the number density was replaced by the mass density to better align with experimental constraints \cite{piscicchia2017csl}, making the GRW collapse rate mass-proportional \cite{pearle1994bound}. However, the stochastic field responsible for irreversibility in CSL has never been connected to gravity.

Both DP and CSL models are non-relativistic, hence by spacetime uncertainties we mean a stochastic fluctuating component of the Newtonian potential. While heuristic extensions of these models towards relativistic regimes are being explored \cite{diosi1990relativistic, ghirardi1990relativistic, pearle1999relativistic}, a fully relativistic theory is still missing. Recently, a general relativistic postquantum theory of gravity has been proposed \cite{oppenheim2023postquantum}, though its renormalizability remains an open question \cite{diosi2024classical}. In our work, we explore a novel relativistic effect. According to general relativity, an uncertainty in the Newtonian potential leads to an uncertainty in the flow of time. We study for the first time the impact of such an effect on the time measured by clocks, as predicted by both the DP and the CSL models. Our analysis shows that this contribution is negligible for any currently available clock. Therefore, SCMs do not place practical limitations on the present precision of time measurements. 

\textit{Spontaneous collapse models}$-$Standard quantum mechanics incorporates two incompatible dynamical principles: a linear, deterministic evolution that enables superposition of states and a non-unitary, measurement-induced process responsible for wavefunction reduction. SCMs provide a unified evolution by modifying the standard dynamical equations with an additional non-unitary term, which becomes significant only for sufficiently macroscopic systems, leading to the quantum–to-classical transition. A particularly compelling class of collapse models employs the mass density operator $\hat{\mu}(\bb{x})$ as the collapse operator, ensuring that large (macroscopic) quantum fluctuations of the mass density are spontaneously suppressed.  

In terms of the density operator $\hat{\rho}(t)$, the dynamics of a system under spontaneous collapse follows the Lindblad master equation
\begin{equation}\label{eq:general ME}
\begin{split}
    \frac{d\hat{\rho}}{dt} =& -\frac{i}{\hbar} [\hat{H},\hat{\rho}] \\ 
    &- \frac{1}{2\hbar^2} \int d^3xd^3y \mathcal{D}(\bb{x}-\bb{y}) [\hat{\mu}(\bb{x}),[\hat{\mu}(\bb{y}),\hat{\rho}]] ,
\end{split}
\end{equation}
where $\hat{H}$ is the standard quantum Hamiltonian of the system and $\mathcal{D}$ is the spatial correlation function of spontaneous $\hat\mu$-measurements at different locations. This correlation function necessarily includes a characteristic smearing (or cut-off) length $\sigma$, which defines the finite spatial resolution of the $\hat\mu$-measurements. 

It is sometimes convenient to express Eq. \eqref{eq:general ME} in an equivalent form using the smeared mass density operator $\hat{\mu}_\sigma(\bb{x}) = \int d^3y g_\sigma(\bb{x}-\bb{y})\hat{\mu}(\bb{y}):= (g_\sigma * \hat{\mu})(\bb{x})$ and a new correlation function $D$ which does not include the smearing, related to $\mathcal{D}$ through $\mathcal{D}(\bb x) = (g_\sigma * D * g_\sigma)(\bb x)$. Here, $*$ is the convolution operator and $g_\sigma$ is a Gaussian function centered at zero with width $\sigma$.

Remarkably, the master equation \eqref{eq:general ME} admits an interesting interpretation in terms of spacetime fluctuations. Using Itô calculus, it can be derived from the following stochastic Schr\"odinger equation 
\begin{equation}\label{eq: stochastic Schroedinger equation}
    \frac{d}{dt}\ket{\psi_t} = -\frac{i}{\hbar} \left[ \hat{H} + \int d^3x \hat{\mu}(\bb{x}) \phi(\bb{x},t) \right] \ket{\psi_t} ,
\end{equation}
where $\phi$ is a classical Gaussian white noise field with zero average and correlation 
\begin{equation}\label{eq: Generic gravity correlation}
    \mathbb{E}[\phi(\bb{x},t),\phi(\bb{y},t')] = \mathcal{D}(\bb{x}-\bb{y}) \delta(t-t') .
\end{equation} 
Eq. \eqref{eq: stochastic Schroedinger equation} corresponds to the ordinary Schr\"odinger equation in the presence of a classical stochastic gravitational field, where $\phi$ plays the role of the Newtonian potential and $\mathcal{D}$ in Eq. \eqref{eq:general ME} is its spatial correlation function. 

Among all the possible SCMs, CSL and DP are by far the most extensively studied. In CSL, the noise is defined by the following unsmeared correlation function
\begin{equation}\label{eq:CSL noise}
    D_\text{CSL}(\bb{x}-\bb{y}) = \frac{\hbar^2 \gamma}{m_0^2} \delta(\bb{x}-\bb{y}) , 
\end{equation}
where $m_0$ is a reference mass, typically chosen to be the nucleon mass. Thus, CSL is characterized by two parameters: $\gamma$, which sets the strength of the collapse process, and the smearing length, usually denoted by $r_C$. It is also customary to use the parameter $\alpha=r_C^{-2}$ as an alternative to the smearing length. Together, these parameters determine the collapse rate for a microscopic system, given by $\lambda = \gamma / (4\pi\sigma^2)^{3/2}$.  The commonly adopted values for the collapse rate and spatial resolution are $\lambda = 10^{-16} \text{ s}^{-1}$ and $\sigma = 10^{-7} \text{ m}$, as proposed by Ghirardi, Rimini and Weber \cite{ghirardi1986unified}.
These values are consistent with most phenomenological analyses of the mass-proportional formulation of the CSL \cite{donadi2021novel,MajoranaBounds}, with the exception of \cite{piscicchia2023novel} which excludes $\sigma$ values smaller than $2\times10^{-7}$ m.

Unlike the CSL, the DP noise strength is set by the Newton gravitational constant $G$ and involves no new parameter. In fact, it is characterized by the following unsmeared correlation function
\begin{equation}\label{eq:DP noise}
    D_\text{DP}(\bb{x}-\bb{y}) = \frac{\hbar G}{|\bb{x}-\bb{y}|} .
\end{equation}

A commonly used reference value for the smearing length, sometimes denoted by $R_0$, is $\sigma = 10^{-9}\text{ m}$. This is approximately five times larger than the strongest lower bound available to date \cite{Lisa2017,donadi2021underground,Adler2019,MajoranaBounds}. 

\textit{Time uncertainty}$-$Building on the previous discussion, we assume a gravitational origin for SCMs attributed to a fluctuating component of the Newtonian potential and explore a relativistic effect associated with such models. We emphasize that the following analysis holds even if we are dealing with non-relativistic SCMs.

In the presence of a classical fluctuating Newtonian field, general relativity requires that time, as well, must exhibit a certain degree of uncertainty. For simplicity, let us assume a flat background spacetime. The fluctuation $\delta t$ measured by a clock at location $\bb{x}$ is given by the relation \cite{diosi2005intrinsic}
\begin{equation}\label{eq: sharp time fluct}
    \delta t(\bb{x},t) = \frac{1}{c^{2}} \int_0^t \phi(\bb{x},\tau) d\tau,
\end{equation}
which is valid in the perturbative regime, where the $00$-metric component is expressed as $g_{00}(\bb{x},t) = 1 + 2\phi(\bb{x},t)/c^2$. Again, $\phi$ is characterized by zero average and a correlation function as given in Eq. \eqref{eq: Generic gravity correlation}. It follows that 
\begin{equation}\label{eq: sharp time correlation}
    \mean[\delta t] = 0, \qquad \mean[\delta t(\bb{x},t)\delta t(\bb{y},t)] = \frac{1}{c^4} \mathcal{D}(\bb x - \bb y) \,t .
\end{equation}
In the following, we characterize the resulting uncertainty associated with any measurement of time. What is the time measured by a clock along its world-tube in the presence of the stochastic noise? If the clock occupies a volume $\mathcal{V}$ in space, the fluctuations will be given by averaging over the volume of the clock the local expression provided in Eq. \eqref{eq: sharp time fluct}. The spatial average of the time correlation function writes as $\braket{ \delta t^2}_\mathcal{V} = \tau t$, where $\tau$ sets the strength of the fluctuation. This quantity has the dimension of time and is defined in terms of the smeared correlation function as 
\begin{equation}\label{eq: general smeared time fluct}
    \tau = \frac{1}{\mathcal{V}^2} \int_{\mathcal{V}} d^3x \int_{\mathcal{V}} d^3y  \frac{1}{c^4}\mathcal{D}(\bb{x}-\bb{y}) .
\end{equation}
The explicit expression of the smeared correlation function $\mathcal{D}$ for CSL and DP models is obtained from the relation $\mathcal{D}(\bb{x}) = (g_\sigma * D * g_\sigma)(\bb{x})$, using Eqs. \eqref{eq:CSL noise} and \eqref{eq:DP noise}, respectively. In the case of CSL, the calculation is straightforward and yields
\begin{equation}\label{eq: CSL correlation}
    \mathcal{D}_\text{CSL}(x) = \frac{\hbar^2 \lambda}{m_0^2} e^{ \frac{-x^2}{4\sigma^2}  },
\end{equation}
where $x=|\bb {x}|$. For DP, it is convenient to perform the calculation in Fourier space, where the convolution is easier to compute. One immediately gets the Fourier transform
\begin{equation}
    \tilde{\mathcal{D}}(k) =  \frac{4\pi\hbar G}{k^2}  e^{-\sigma^2k^2} ,
\end{equation}
from which the smeared spatial correlation function of the gravitational potential is derived as 
\begin{equation}\label{eq: DP correlation}
    \mathcal{D}_\text{DP}(x) = \frac{\hbar G}{x} \text{erf}\left( \frac{x}{2\sigma} \right) .
\end{equation}

\begin{figure}[b]
  \centering
  \includegraphics[width=1\linewidth]{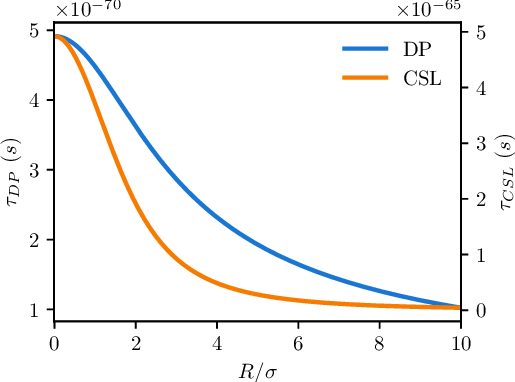}
  \caption{\justifying Time fluctuation strength as a function of the ratio between the clock's radius and the smearing length. The DP case (blue) is plotted on the left axis, and the CSL case (orange) is plotted on the right axis. The smearing length and the collapse rate parameters are set to the standard values $\lambda = 10^{-16}$ s$^{-1}$, $\sigma_\text{CSL}=10^{-7}$ m and $\sigma_\text{DP}=10^{-9}$ m.}
  \label{fig: time fluctuation}
\end{figure}
Then, the evaluation of $\tau$ can be carried out analytically by choosing, for instance, a spherical shape for the clock. The exact results are shown in Fig. \ref{fig: time fluctuation} for a specific range of the clock radius $R$ and a choice of parameters corresponding to the standard reference values. For both CSL and DP, the fluctuation in time measurement is maximal for clock sizes comparable to or smaller than the smearing length $\sigma$. In contrast, larger clocks become less sensitive to the uncertainty of spacetime as $R$ increases, leading to more precise measurements of the mean time. This is because the clock measures the mean time within the section $\mathcal{V}$ of its world-tube. The optimal spatial resolution corresponds to the smallest size of $\mathcal{V}$, when time-keeping has the worst precision. Conversely, as $\mathcal{V}$ increases, time-keeping precision improves at the expense of spatial resolution. The asymptotic behavior of the fluctuation strength is as follows
\begin{align}\label{eq: small clocks CSL}
    &\tau_\text{CSL} \sim  \frac{\hbar^2 \lambda}{m_0^2c^4} =: \tau^\text{max}_\text{CSL} , \\[0.2cm]
    & \tau_\text{DP} \sim  \frac{\hbar G}{\sqrt{\pi}c^4\sigma } =: \tau^\text{max}_\text{DP}\label{eq: small clocks DP} ,
\end{align}
for  $R\lesssim\sigma$ and
\begin{align}\label{eq: big clocks}
    &\tau_\text{CSL} \sim \frac{6\sqrt{\pi}\tau^\text{max}_\text{CSL} }{(R/\sigma)^3} , \\[0.2cm]
    &\tau_{DP} \sim \frac{6 \sqrt{\pi}\tau^\text{max}_\text{DP} }{5 (R/\sigma)} ,
\end{align}
for $R\gg\sigma$. 

In conclusion, the relativistic relation \eqref{eq: sharp time fluct} combined with the presence of a stochastic component in the gravitational field implies an unavoidable uncertainty in any time measurement. Regardless of how small $\tau$ may be, if it is nonzero, this uncertainty inevitably increases over time. Therefore, it is essential to consider whether this uncertainty could lead to measurable effects, potentially imposing a fundamental limit on clock precision.

\begin{figure}[b]
    \centering
    \includegraphics[width=1\linewidth]{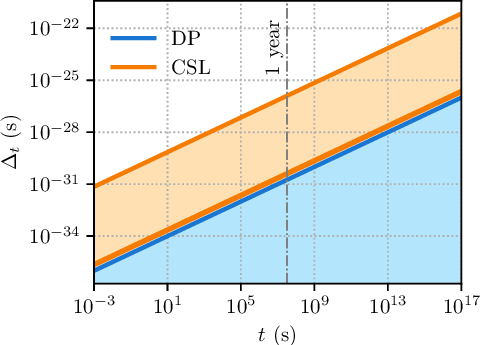}
    \caption{\justifying Time uncertainty for optimal clocks as a function of time, extending up to the age of the universe. The colored areas correspond to the experimentally allowed regions. The blue region encompasses all allowed values constrained by the lower bound on the DP smearing length, $\sigma_\text{DP} = 4.94\times 10^{-10}$ m. The orange region corresponds to the range permitted by experimental bounds on the CSL collapse rate, $10^{-20}\ \text{s}^{-1} < \lambda < 10^{-11}\ \text{s}^{-1}$. The upper edges of both regions represent the maximum time uncertainty predicted by the DP and CSL models.}
    \label{fig: uncertainty}
\end{figure}

After a period of time $t$, the random component of the gravitational field induces a fluctuation in the measured time given by $\Delta_t :=\sqrt{\braket{ \delta t^2}_\mathcal{V}} = \sqrt{\tau t}$. We focus on optimal clocks with respect to spatial resolution, defined as those with dimensions comparable to the smearing length, for which the fluctuation strength reaches its maximum value $\tau^\text{max}$ (Eqs. \eqref{eq: small clocks CSL} and \eqref{eq: small clocks DP}). To provide insight into the relevant orders of magnitude, we set the collapse parameters to their reference values and find, at $t=1$ year, the values
\begin{equation}\label{eq: time fluctuation with standard values}
    \Delta_t \simeq 
    \begin{cases}
        10^{-28}\, \text{s in 1 year for CSL} \\[0.2cm]
        10^{-31}\, \text{s in 1 year for DP} 
    \end{cases} 
\end{equation}
For the CSL case, the fluctuation strength is proportional to the microscopic collapse rate. Current experimental constraints on this parameter, reported in \cite{MajoranaBounds}, are approximately given by $10^{-20}\ \text{s}^{-1} < \lambda < 10^{-11}\ \text{s}^{-1}$. Correspondingly, for these values, the time fluctuation $\Delta_t$ at $t=$ 1 year ranges from $10^{-31}$ s to $10^{-26}$ s. On the other hand, for what regards the DP model, $\tau^\text{max}$ is inversely proportional to the smearing length. Experimental bounds place a lower limit on the latter at  $4.94\times 10^{-10}$ m \cite{MajoranaBounds}. Fig. \ref{fig: uncertainty} shows the uncertainty affecting optimal clocks corresponding to all the experimentally allowed values of the CSL and DP parameters, from which one can see the maximum uncertainties predicted by the CSL and DP models.   

We now compare these results with the highest precision achieved by modern clocks. All time-keeping devices operate referencing a stable oscillatory phenomenon, whether it is the swing of a pendulum, the periodic motion of celestial bodies or the vibration of a quartz crystal. Fluctuations in time measurements due to noise processes are then quantified by the fractional stability $\sigma_y(t)$, which characterizes the frequency stability of the clock over a given period $t$. The corresponding fluctuation in the measured time is given by $\Delta_t = \sigma_y(t) t / \sqrt{3}$. 

Atomic clocks - and potentially future nuclear clocks \cite{zhang2024frequency} - are the most precise devices, relying on the exceptional frequency stability of the electromagnetic radiation emitted during atomic transitions. Among these, optically trapped neutral-atom clocks currently set the benchmark for frequency stability, leveraging large ensembles of atoms to reduce statistical noise beyond the limits of single-ion clocks \cite{PhysRevLett.134.023201}. State-of-the-art optical lattice clocks based on strontium or ytterbium atoms achieve short-term fractional frequency stabilities on the order of $10^{-17} / \sqrt{t/1\text{s}}$ \cite{oelker2019demonstration,schioppo2017ultrastable,bothwell2022resolving}, corresponding to time fluctuations of approximately $10^{-17}\text{ s} \cdot \sqrt{t/1\text{s}}$ over periods of hours or days. 

Over longer time intervals (years or decades), clocks based on millisecond pulsars can approach frequency stabilities comparable to those of atomic clocks \cite{hobbs2020pulsar, rodin2022generalized, dong2017ensemble}. However, pulsar clocks are expected to be insensitive to noise induced by SCMs, as this would be suppressed by the extremely small ratio $\sigma/R$. Consequently, they could serve as a tool for measuring this type of uncertainty in atomic clocks. 

Nevertheless, given these capabilities, it is evident that the uncertainty associated with SCMs is entirely negligible compared to the performance of current time-keeping technology. 

\textit{Concluding remarks}$-$In this letter, we addressed the implications of spacetime uncertainty within the context of spontaneous collapse models, focusing on their mass-proportional versions. In this class of models, which includes the leading CSL and DP proposals, the wave function collapse dynamics can be attributed to the continuous and spontaneous monitoring of the mass density operator.

We highlighted that the stochastic field defining these models can be interpreted as the Newtonian gravitational potential, possibly relating this class of models to gravity. As a direct consequence, fluctuations in the gravitational field induce an intrinsic uncertainty in the flow of time. We emphasize that while our analyses rely on non-relativistic collapse models, this effect is fundamentally relativistic in nature. For the first time, we investigated the impact of this time uncertainty on clock measurements. Our results reveal that the effect is maximized for clocks with sizes comparable to the spatial resolution of the collapse (the smearing length $\sigma$), while for larger clocks the fluctuations average out, enabling them to measure mean time with higher precision - at the price of lower spatial resolution. Moreover, irrespective of the fluctuation strength, the uncertainty grows unavoidably over time.

We conclude that while spontaneous collapse models introduce a fundamental source of time uncertainty, they do not impose any practical limitation on precision time-keeping. Our findings offer a new perspective on the interplay between quantum collapse models and gravitational effects, paving the way for further exploration of their possible observational signatures. \\

\textit{Acknowledgments}$-$This publication was made possible through the support of Grant 62099 from the John Templeton Foundation. The opinions expressed in this publication are those of the authors and do not necessarily reflect the views of the John Templeton Foundation. We acknowledge support from the Foundational Questions Institute and Fetzer Franklin Fund, a donor advised fund of Silicon Valley Community Foundation (Grants No. FQXi-RFP-CPW-2008 and FQXi-MGB-2011), and from the INFN (VIP). D.L. was supported by the National Research, Development and Innovation Office “Frontline” Research Excellence Program (Grant No. KKP133827). C.C. and D.L. benefited from the EU COST Actions CA23115 and CA23130. N.B. and K.P. acknowledge support from the Centro Ricerche Enrico Fermi - Museo Storico della Fisica e Centro Studi e Ricerche “Enrico Fermi” (Open Problems in Quantum Mechanics project).

\bibliography{bib}

\end{document}